\documentclass[conference]{IEEEtran}
\IEEEoverridecommandlockouts
\usepackage{cite}
\usepackage{amsmath,amssymb,amsfonts}
\usepackage{algorithmic}
\usepackage{graphicx}
\usepackage{textcomp}
\usepackage{xcolor}
\def\BibTeX{{\rm B\kern-.05em{\sc i\kern-.025em b}\kern-.08em
    T\kern-.1667em\lower.7ex\hbox{E}\kern-.125emX}}
\begin{document}

\title{Automatic Pipeline Provisioning\\
}

\author{\IEEEauthorblockN{Alexandre-Xavier Labonté-Lamoureux}
\IEEEauthorblockA{\textit{Département de Génie Logiciel} \\
\textit{École de Technologie Supérieure}\\
Montréal, Canada \\
alexandre-xavier.labonte-lamoureux.1@ens.etsmtl.ca}
\and
\IEEEauthorblockN{Simon Boyer}
\IEEEauthorblockA{\textit{Département de Génie Logiciel} \\
\textit{École de Technologie Supérieure}\\
Montréal, Canada \\
simon.boyer.1@ens.etsmtl.ca}
}

\maketitle

\begin{abstract}
The goal of this paper is to explore the benefits of automatic pipeline provisioning and identify how it can be applied. Automatic pipeline provisioning can be defined as a process of quickly deploying a pipeline for a software engineering project. This research will focus on CI pipelines, although the outcomes of this approach on CD pipelines will likely be similar. 
\end{abstract}


\section{Introduction}
Automation is a central part of DevOps. Developers will create CI pipelines to automatically compile and test their software as new code changes are uploaded to a source-control management system. Despite the widespread adoption of automation in DevOps, there is a specific area that has not been fully explored: automatic pipeline provisioning. This process involves the rapid deployment of a CI pipeline for a software project. This automation can be used to significantly streamline the creation of CI pipelines. 

The focus of this research is Continuous Integration (CI) pipelines. A CI is a development practice where developers frequently integrate code into a shared repository. They preferably do this several times a day. Each new integration is verified by an automated build system and, preferably, automated tests are also performed. There are well know benefits to this such as allowed for a rapid development pace, but the potential advantages of automatic pipeline provisioning are not yet well understood. 

This paper aims to demonstrate the benefits of automatic pipeline provisioning and identify its shortcomings. It will also explore how it can be effectively implemented in a CI environment. Although this research will focus on CI pipelines, it is reasonable to assume that the findings would be applicable to Continuous Deployment (CD) pipelines given that there are similarities between both. 

In the following sections, the authors will delve deeper into what automatic pipeline provisioning entails, its potential benefits, its different kind and levels of implementation, and finally a case study will be presented. The explanation of automatic pipeline provisioning in the context of CI found in this paper will provide valuable insights for organizations looking to optimize their DevOps practices.

\section{Problem statement}
CI pipelines are now ubiquitous in the development pipeline: 71.93\% of developers reported having CI tooling available at their organization according to the 2023 Stack Overflow Developer Survey\cite{b1}. Some say having to configure a CI pipeline for every project can become a bane for developers as it can be error-prone, in addition to being time-consuming and a repetitive task \cite{b2}. A CI pipeline must be created if a developer wants to ensure that every commit pushed to the source-control management system is compilable, passes the unit tests and security tests in an automated manner. Time spent on creating and maintaining the CI script is time lost as it is not time spend on developer new features that bring value to the customer. In a paper on DevOps tools diversity, Mik Kesrten explains that "as software development becomes more complex, so will the specialization of the toolchain" \cite{b3}. Because of this, it becomes more and more difficult for developers to develop and maintain their increasingly complex pipelines. Furthermore, if the developers were to migrate to a new CI infrastructure, the CI scripts must be completely rewritten which is time consuming. This hefty time cost may prevent developers from opting to migrate to another platform, which can constraint organization change in the long run.

\section{A better approach to CI pipelines}
A new approach to the creation of CI pipelines would be to automate their creation. As with the automation of many tasks in the software life cycle, which is a core principle of the DevOps philosophy, developers should extend this automation to non-creative tasks like CI pipelines. 

As to further improve the pipelines, projects should use certain already ascertained golden paths. Golden paths are a limited set of paths that already implement the best practices of software engineering and, as Gary Niemen says on the Spotify blog, they are opinionated and are workflows the organization is ready to support\cite{b4}. This practices ensures a high maturity level is kept among all projects in the organization. It is usually the most senior employees in the organization who will create these paths as they are the most experienced and can foresee long term issues that may arise in certain paths. There are challenged to maintaining these paths, as a change to a CI script that is used as a golden path may break the build process of some projects as it will be reused across these projects. 

In an organization, golden paths will be used as templates, which are part of codes that can be reused. By doing this, developers avoid rewritting new CI scripts for projects or copy-pasting. If a CI script is not templated and rather copied, an improvement implemented in a CI script is not automatically carried out in the other scripts. This lead to a diversion between CI scripts and, over the long run, improvements will require a larger effort to port into the other CI scripts. When configuration significantly diverge from each other, this is called a “configuration drift” \cite{b5}.

Developers should not be given access to the CI scripts which they usually have access to when they have access to a project. Developers would have a tendency to edit their local project CI script to add something they want for what they perceive as an immediate improvement. This does not benefit them in the long run. 



\section{Feasibility}
In an organization which has not already moved towards templated CI scripts, especially a large one, there will likely be a different CI script in every software project. The organization should establish a committee that will evaluate CI scripts currently used in the organization and use them as a basis for creating golden paths. The committee will then pick a set of projects to be used in a proof of concept so they can see, over time, how this set of golden paths is used and what value they get from it. 

Golden paths must exist for each language used in the organization. If there are many tools which accomplish the same goal in the CI pipelines, for example Trivy and Snyk or Deep Source and SonarQube, they should be consolidated. This will avoid duplicate effort and increase efficiency. The point of a golden path is to always use the best tools in every scenario. Liberty is reduced in favor of standardization and automation. These practices, while sometimes constraining, will lead to greater efficiency, consistency, and reliability in technology organizations. They might not feel constraining at all for the developers as not having to think about this could result in a reduction of mental load. 

\section{Implementation of this approach}
There are a few ways to implement automatic CI pipeline provisioning. They will differ according to the complexity of the implementation that is sought and the platform being used, whether it is Bitbucket, GitHub or GitLab.

\subsection{GitLab CI}
\begin{itemize}
\item Import a template of the CI into the software project. It cannot be customized in any way and its behavior cannot be changed.
\item Import a template of the CI into the software project, but this template is configured via environment variables.
\item Import templates of jobs that the developer wants to use. Each job can also be configured via environment variables.
\item The CI script can be generated from a template engine ahead of time. The template engine always runs in the first step of the CI script to verify that the script has indeed been generated by itself and has not been modified by the developer before continuing with the next jobs in the CI script. A different implementation could leverage downstream pipelines to simplify this process. It allows passing a custom CI/CD configuration file that was generated on the fly to another pipeline.
\end{itemize}

\subsection{GitHub Actions}
\begin{itemize}
\item A limited set of Actions can be made available to the developers.
\item The use of the "run" keyword in a workflow has been disabled. This constrains developers to existing or available GitHub Actions.
\item A reusable workflow which selects the right jobs to be run according to the content of the source code repository could be programmed using GitHub Actions' powerful TypeScript capabilities.
\end{itemize}

\subsection{Others}
\begin{itemize}
\item The CI template engine can be integrated into the CI/CD engine of the platform of choice. Not every CI/CD engine will support this, but Concourse CI can call a program when the CI pipeline is meant to be run to generate all the CI steps for a project at run-time. This is achieved through a \textit{set\_pipeline} step. Through changes to the template engine, the organization's Platform team, where the design of CI pipelines is centralized, can control what is any of the organization's pipelines. This can be leveraged in many ways, such as quickly deploying third-party dependency scanning when a vulnerability like Log4j is discovered for example.  
\end{itemize}

\section{Implementation on GitLab CI}
In this section, different strategies aforementioned will be explored using GitLab CI. 

\subsection{Template repository}
For this section, the usage of a template directory on GitLab CI will be demonstrated. 

First, the repository must be created. The \textit{readme} file of the repository should contain usage information and document the structure of the repository.

Second, it must be populated with a hierarchy of subdirectories to keep things organized. There are many ways to organize them. The directories can versioned, meaning that the root directory is a version number. This way, breaking changes can be introduced without breaking the projects' CI. Another approach may be to use branches as versions, where the name of the branch is the version string. Developers might elect to do rolling updates in order to decrease maintenance, but this will increase the risk of breaking pipelines if done incorrectly.

Third, the directories must be populated with reusable pipeline jobs. For instance, a reusable Pylint job should be found inside the \textit{test} directory of a \textit{python} directory. A tool compatible with many programming languages, such as Trivy (a security scanner) should be put under the \textit{sast} directory. A build job for the Go programming language should be found in the \textit{build} directory inside a directory named \textit{go}. Developers might configure the Go version used using an environment variable set in their CI job or the specific YAML template should have the Go version in its file name. Someone could think it is a great idea to always use the latest version of Go as it is backwards compatible. Issues may arise when Protocol Buffer generated code has be committed to the directory. This code is version-specific and will differ when generated using another version of the Go compile. Protocol Buffer are very popular with microservices, but an inexperienced programmer might not be aware of that. This is why most decisions regarding the architecture of the template directory should be taken by the most senior software developers in the organization. Due to experience, they are more inclined to be able to predict and see problems before they arise. 

Any implementation of this does not have to follow this structure, members of an organization will find what is best for them through experimentation. What has been described here is what was used as a structure in the example project. 

As to better illustrate what is being talked about, an example project has been created as seen in Fig. 1. The project is named \textit{Pipeline Blocks} as the principle here to use blocks of pipeline code like Lego bricks to build a complete CI pipeline. 

\begin{figure}[htbp]
\centerline{\includegraphics[scale=0.2]{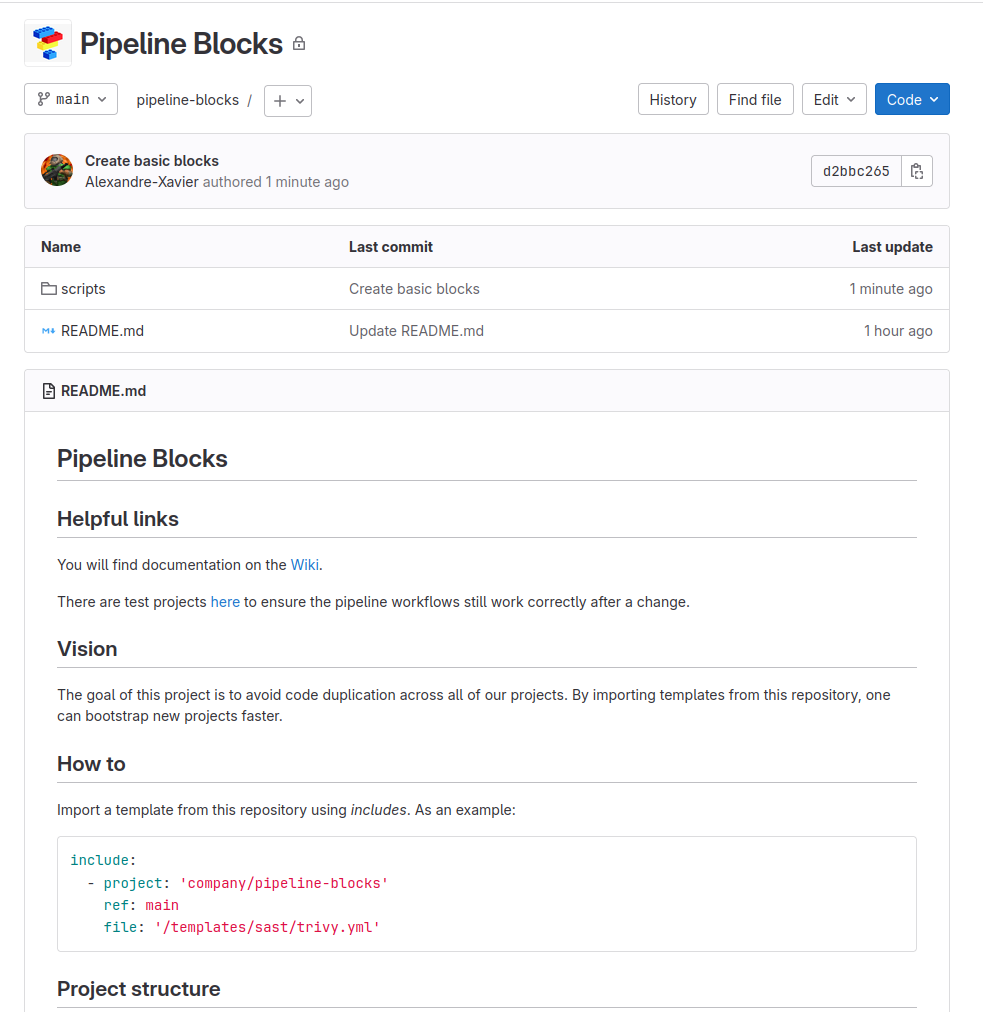}}
\caption{GitLab repository of the Pipeline Blocks project}
\label{fig}
\end{figure}

The directory structure is show in the following figure. There are currently two programming languages supported: Go and Python. Markdown files such as \textit{readme} files are omitted from Fig. 2. 

\begin{figure}[htbp]
\centerline{\includegraphics[scale=1.0]{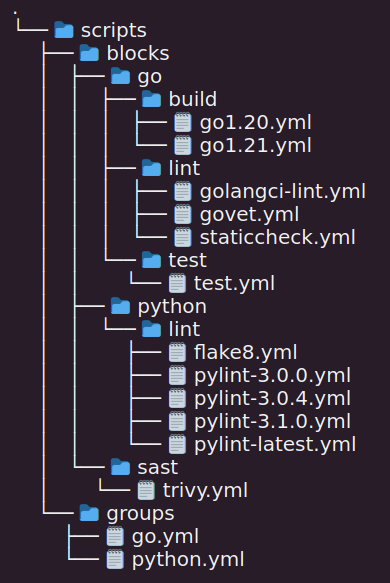}}
\caption{Directory structure of Pipeline Blocks}
\label{fig}
\end{figure}

The golden paths that are supported can be seen in the directory structure. For Go, there two versions of the compiler currently supported. A few linters which were vetted by the most seniors developers are currently offered. There is only one possible test job offered, but it runs the \textit{make test} command which the developers can program to do anything in their makefile. 

In the case of Python, two different linters are offered. The developers can pick one of the three versions of Pylint or they may opt to use the latest version to always stay up-to-date. Each version comes with a set of rules that was improved by the organization's most senior Python programmers. This set of rules is included in the job and will be updated in this single location. This allows every project in the organization to run the latest set of rules if they use this job from the Pipeline Blocks project. 

Another job is provided for security. This one is shown as an example in Fig. 3. It currently is set up so the scanner will return an error code in the pipeline if the security of a project does not meet the organization's standards. If the security team would like to improve the organization's cybersecurity standards, they could simply edit this single files to make a change to every project in the organization (assuming every project in the organization includes this file, which they should because they follow the enforced best practices.)

\begin{figure}[htbp]
\centerline{\includegraphics[scale=0.14]{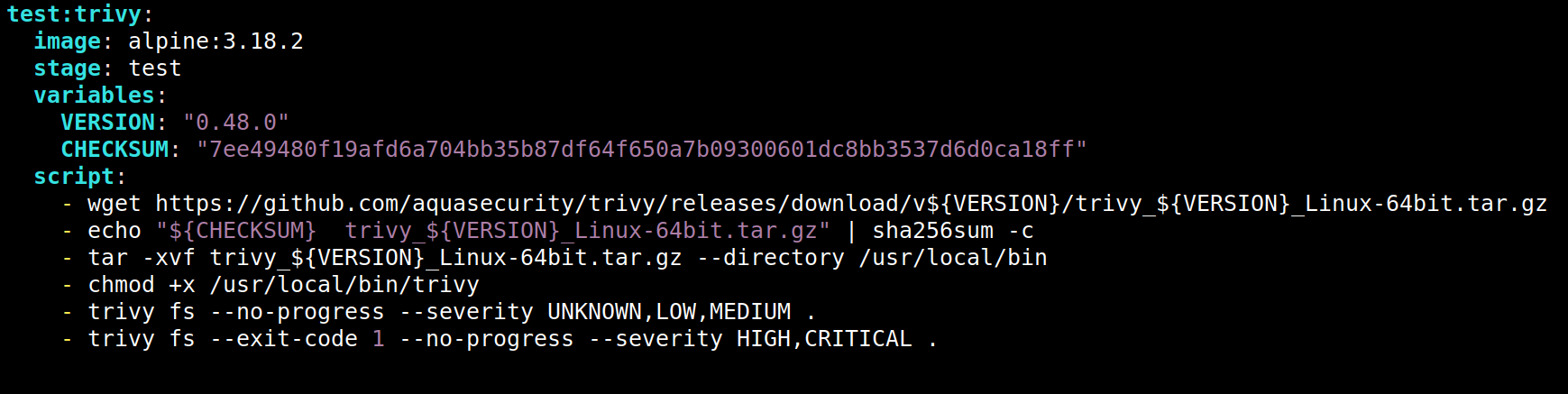}}
\caption{Content of the \textit{trivy.yml} template}
\label{fig}
\end{figure}

The Trivy security scanner is recommended by the organization's cybersecurity team. As such, if the developer chooses to use \textit{groups}, the Trivy scanner will run in the project's pipeline. \textit{Groups} are a collection golden paths for a single programming language. Thus, in the case of a project written in Go, a developer may include the \textit{go.yml} file that is in the \textit{groups} directory. This will cause the user to have one of the build jobs, all the linter jobs and the test job in their project's pipeline. If the developers didn't include the \textit{groups}' Go template, they would have to include each job of their choice manually into their GitLab CI pipeline file. 

\pagebreak 

\begin{figure}[htbp]
\centerline{\includegraphics[scale=0.225]{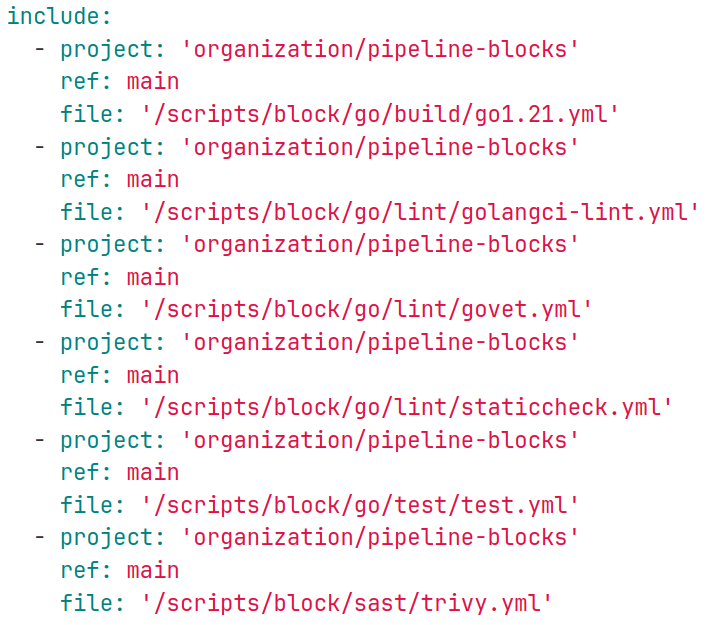}}
\caption{The file \textit{go.yml} in the \textit{groups} directory. The developers only have to include this file to include all the other pipeline jobs.}
\label{fig}
\end{figure}

Upon including the template file for the group in their \textit{gitlab-ci.yml} file, the developers have a full CI pipeline generated for them. The group template file is shown in Fig. 4 and the resulting CI pipeline is shown in Fig. 5.

\begin{figure}[htbp]
\centerline{\includegraphics[scale=0.25]{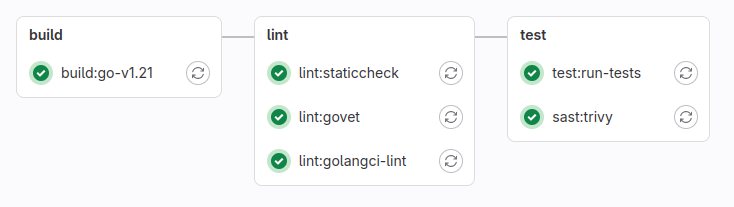}}
\caption{Resulting CI pipeline in the example project}
\label{fig}
\end{figure}

This is an effortless manner of managing pipelines and is scalable. An update to a templated job can easily be propagated to every project in the organization. The benefits of this have yet to be measured. This is often called a cookie cutter approach and its goal is to improve the developer experience\cite{b6}. 

\subsection{GUI-driven CI provisioning}
Instead of configuring a project via YAML code, some might prefer using a GUI to set up a project (Fig. 6.) In the next section, you will read about Internal Developer Portals which follow the same principle, but are as much or sometimes more powerful. 

\begin{figure}[htbp]
\centerline{\includegraphics[scale=1.25]{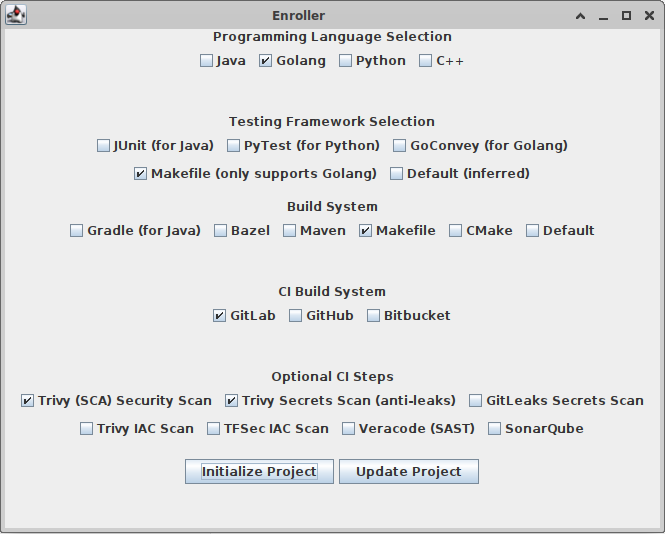}}
\caption{GUI used to setup a project and generate a basic CI pipeline}
\label{fig}
\end{figure}

Using a GUI is less rigorous, as it is not integrated directly into the CI in any sort of way. Thus, the software will not control when the CI pipeline script is updated. Its purpose is to generate a new CI file or update the current one. If the maintainers of this tool want to push an update to every project in the organization, the project owners will have to launch the GUI and regenerate the CI by themselves. 

It offers a certain level of protection to the CI script from manual changes in the sense that it abstracts it and the developer does not have to see the script in order to make changes. Every supported functionality or golden path is built into this software. The GUI is an intermediary that provides an abstraction layer over the CI pipeline which may not even be formatted to be human readable as it is not its purpose. Any manual changes would be lost on the next run of the software, as such it does not encourage any change that is outside of the supported golden paths.

Per se, this is a project bootstrapping or project onboarding software. This concept can be pushed further, and the same type of tooling could be created for infrastructure provisioning. Such a tool would generate Infrastructure as Code (IaC) that could immediately be used to provision the cloud infrastructure required for a software project. IaC is not a subject that will be delved into in this paper, but what is highlighted here is that the same principles can be applied to other parts of the software development lifecycle in order to enhance the developer experience.

\subsection{Fully automatic CI provisioning}
Fully automatic CI provisioning is described as the automatic provisioning of CI pipelines scripts without any intervention from the developers or with only an iota of developer intervention. This is the highest degree of automation someone could get. Either there is a CI template engine can be integrated into the CI/CD engine of the platform that is used, or it runs in the first step of the CI pipeline and will ask the developer to update the repository's CI script only if required. There are a few reasons why the template engine might ask the developer for a manual intervention in order to update the CI script: Either there was a manual change made to the CI script or there is an updated available for the golden paths which the template engine cannot apply by itself due to the nature of the CI/CD engine of the platform used. 

For example, when using GitLab, a \textit{gitlab-ci.yml} file must exist at the root of the source code repository. It represents the state of the pipeline at the current commit. It has to be replaced manually by a developer when they will commit to the repository. A CI template engine can still be used to generate this file, but if any changes are required, the developer has to run the template engine by himself or herself. The template engine has to be integrated in one of the pipeline jobs (likely the first one) to verify that the \textit{gitlab-ci.yml} file is up-to-date before running the CI pipeline. This ensure the CI script is in the expected state and that manual changes were not made. Otherwise, a \textit{downstream pipeline} could be trigger a generated CI script received as an artifact to execute.\cite{b16}

There are other CI/CD engines which support Dynamic Configuration Management. According to a Humanitec blog by Chris Stephenson, DCM is a methodology used to structure the configuration of compute workloads where developers can create workload specifications, describing everything their workloads need to run successfully\cite{b7}. Arguably, it is not necessarily the developer who will create their workload specifications. CI/CD engines like CircleCI or Concourse (via the \textit{set\_pipeline} step) can dynamically create their own CI pipeline from specifications inferred from the project. A CI template engine such as the one described in the previous paragraphs can be inserted here to generate the pipeline steps dynamically. A repository would not need to contain a file such as \textit{.gitlab-ci.yml} on GitLab or files that would usually be found in the \textit{.github/workflows} on GitHub. 

To infer the required pipeline for the project, the CI template engine would scan files in the current Git repository to determine the type of CI pipeline it must create. It will scan the extensions of code files to determine the programming language used in order to generate the correct \textit{build} step. It will also attempt to find a \textit{makefile} file, in which case the \textit{build} step will consist of running the \textit{make build} command. If the programming language is Go, it will search for files ending with \textit{\_test.go} in order to know if it should add a \textit{test} step to the pipeline. It may also attempt to run the \textit{make test} command for other programming languages in order to run the tests. No \textit{test} step is generated if no tests are found. If it finds a \textit{go.mod}, \textit{requirements.txt} or a \textit{pom.xml} file, it will know that it should generate a Software Composition Analysis (SCA) step in order to scan for vulnerabilities in third-party dependencies. If it finds IaC files such are Terraform \textit{.tf} files, it will create a SAST step which may be running TFSec, Checkov or any other misconfiguration scanner. The CI template engine may also generate the CD part of the pipeline with a step that runs a command that will deploy the IaC code on approval according to the organization's policies and which may automatically retrieve the secrets required to deploy from the secure store such as HashiCorp Vault with the developer having no idea, and no need to learn, how to interact with Vault.

\section{Internal Developer Portals}
As DevOps and CI/CD tooling evolves and diversify, some new tools that acts as an abstraction for it are getting more visibility: Internal Developer Portals. Examples of such tools are Backstage (backstage.io) and Port (getport.io). Integrating automatic pipeline provisioning with portals allows a simple interface to setup everything needed. For this paper, Port (getport.io) will be used to demonstrate how to leverage portals for automatic pipeline provisioning. 

\subsection{Data model}

Port uses data models to define relationships between entities, here there are four entity types as illustrated in Fig. 7:
\begin{enumerate}
    \item \textbf{Application:} The application is a high-level entity defined by the user. It contains all requirements for an application (composed of one or more repositories), including test, security and build requirements.
    \item \textbf{Repository:} Repositories are imported from the version control system and are then linked to an application. The user can then select languages and toolchains for that repository, information that will be used to provision CI pipelines.
    \item \textbf{CI Engine:} A list of possible CI engines to provision is defined and can be linked to a repository. This allows the pipeline generation process to adapt the format to different CI Engines.
    \item \textbf{CI Pipeline:} A list of standard pipelines and/or pipeline templates can be defined in Port and linked to repositories by an automatic that select appropriate pipelines based on the application, the repository and the CI engine.
\end{enumerate}

\begin{figure}[htbp]
    \centering
    \includegraphics[width=1\linewidth]{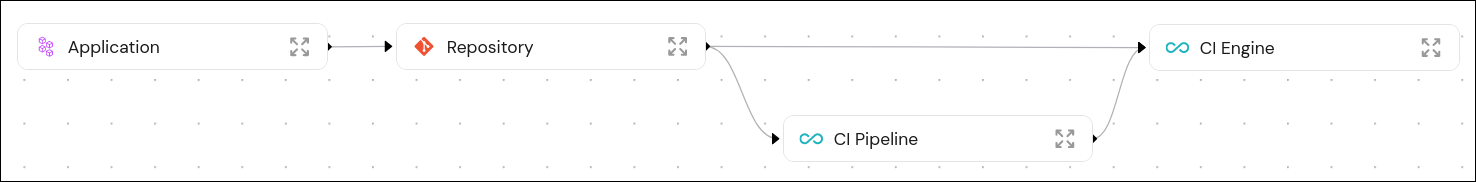}
    \caption{Port Data Model}
    \label{fig:enter-label}
\end{figure}

\subsection{Application and Repository creation}

Repositories are automatically imported in Port, but some information, like the language and the toolchain used, cannot be imported. Using a custom form that will allow a developer to "setup" a repository will add missing context to the data model to make it easier to generate pipelines later as shown in Fig. 8.

\begin{figure}[htbp]
    \centering
    \includegraphics[width=1\linewidth]{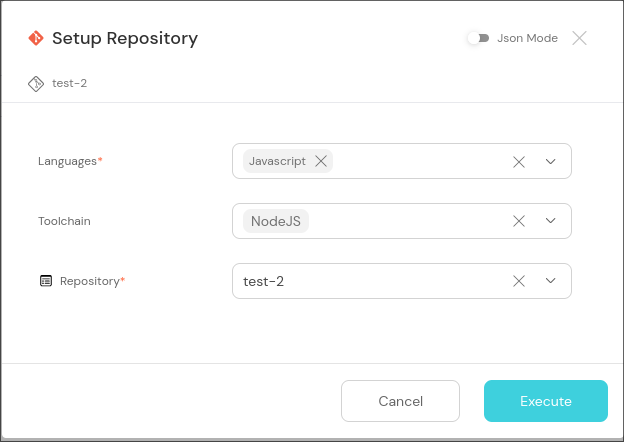}
    \caption{Git repository setup using Port}
    \label{fig:enter-label}
\end{figure}

Additional context is also added by adding a form to create an application and link repositories to it (Fig. 9.) Here, application-wide information can be added, such as if linting is required or the target code coverage. Note that this application-repository aggregation model is not essential to the concept presented here; it merely demonstrates how information can be organized within a developer portal.


\begin{figure}[htbp]
    \centering
    \includegraphics[width=1\linewidth]{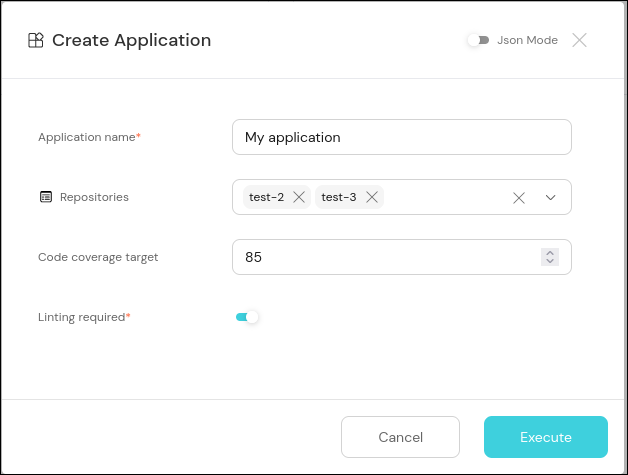}
    \caption{Application creation using Port}
    \label{fig:enter-label}
\end{figure}

\subsection{CI pipeline generation}

To generate pipelines, the previous information from the other entities can be used to select the appropriate pipeline templates. Those templates can be written by people specializing in CI/CD, they are the "golden paths". With a developer portal, a developer can then register them into the platform so that they are selected when the different properties matches.

\begin{figure}[htbp]
    \centering
    \includegraphics[width=1\linewidth]{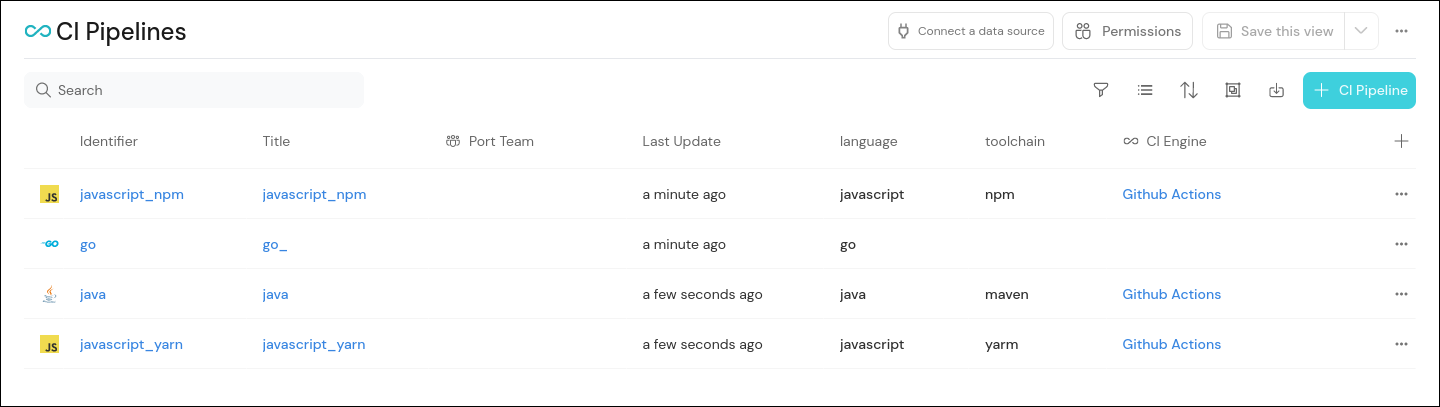}
    \caption{CI pipelines list in Port}
    \label{fig:enter-label}
\end{figure}

\subsection{Result}
Using a developer platform, it is now possible to automate the creation of new pipelines using context information provided about the application and the repository through a simple to use UI. Compared to manual imports of templates, this is easier and reduces the risk of manipulation errors. Keeping an internal data model also helps track which repository are using which templates, which can help when a template is updated. In general, having those information available in the developer portal allows other components of the development cycle, like the deployment, to reuse these same information to create a more consistent and less redundant environment.

\section{Case Study}

To study the potential impacts of automatic pipeline provisioning, Company A will be used as an example. Company A is a financial institution migrating from Jenkins to Github Actions. In this process, they first started with the creation of custom pipelines for specific context. Later on,  they decided to evaluate the potential impacts of using a developer portal to provision pipelines.

This case study will look at a standard pipeline that was developed both without automatic pipeline provisioning and with automatic pipeline provisioning. This pipeline includes the following steps:

\begin{enumerate}
    \item Build a Java application with Maven
    \item Run application unit tests
    \item Run a security scan on the application
    \item Push the package to Nexus
    \item Deploy the application to Azure Websites
\end{enumerate}

\subsection{Without automatic pipeline provisioning}

Company A has a lot of approval processes due to regulatory requirements. Creating a new pipeline brings a lot of challenges as everything needs to be tested and validated by multiple teams to insure that everything respects those regulations.

Because of this, the process of creating a new pipeline often requires a lot of hours. Here, for the process of creating the standard pipeline of this case study, it took a CI/CD expert \textbf{150 hours} spread out on 2 months. Most of this time was spent on organizational approvals. If each team needs to spend this amount of time or more on setting up their pipeline, it represents a significant investment from the company viewpoint.

Company A realized that this case-by-case method will be difficult to scale, particularly since its developers are not CI experts.

\subsection{With automatic pipeline provisioning}

For automatic pipeline provisioning, company A decided to use a mix of templates and reusable components. Everything that required some approvals and are regulatory requirements were put into easy to use reusable workflows:

\begin{enumerate}
    \item \textbf{Validate:} Runs the required security checks and adds a signature to the commit so that it can be pushed.
    \item \textbf{Push:} Pushes a build that passed the \textit{Validate} workflow to Nexus
    \item \textbf{Deploy to Azure Websites:} Deploys only Nexus packages (already validated) to an Azure website.
\end{enumerate}

Using this list of required workflows, it allows to shift some of the complexity around validation, signature and organizational processes away from the developer. With this, the developer still has control over how his application is built, how the unit tests are ran and the ordering of the different steps according to his or her needs. These reusable workflows are not alone an automatic pipeline provisioning strategy, but rather a way of organizing tasks that will help the provisioning.

To automate the actual provisioning of the pipelines, company A chose to use repository templates. To do so, it used some common build and test patterns in its projects and defined repository templates for it. Now, when creating a new repository, the developer can choose the appropriate template. This will combine a build step, a test step and the three reusable workflows in two standard flows: one for a pull-request and one for a merge commit on the head branch.

Configuring all of this for one set of language and tooling took \textbf{200 hours}. Again, most of it was spent on organizational approvals, but because of the reusable workflows and templates, these steps should not be required for the next projects using these templates.

The automatic provisioning itself of a new pipeline now only takes a few minutes.

\subsection{Comparison}

CI pipelines are tightly linked to company processes and requirements, which can make it long and difficult to build one that follows all the company rules. This is particularly the case for company A, where 150 hours were required to build one CI/CD pipeline. By spending a third more time on this (200 hours), this company was able to build a repeatable and automatic pipeline provisioning system, which will be pay off after only two uses of this pipeline provisioning setup. The maintenance being centralized in reusable workflows and templates will also make it easier for company A to maintain its pipelines up to date.

\section{Related work}





The automation and optimization of CI/CD pipelines have begun to be explored in recent research, with a focus on reducing manual effort, minimizing errors, and improving consistency. Below are is a summary of key contributions in this domain.

Employees of ByteDance Ltd. have written a paper \cite{b8} about 
PIPELINEASCODE, the company's own CI/CD workflow management system. Using this system, which has shown great benefits, developers can more easily build CI/CD pipelines. Their research has shown that the new pipelines have a 25\% higher success rate, builds are less flaky, and encourage a 12\% higher change frequency, despite the pipelines being built in PIPELINEASCODE being more complex.  Their system allows building pipelines through a GUI using "atoms" which represents a reusable CI/CD pipeline step.

In the paper \textit{A Model-driven Approach to Continuous Practices for Modern Cloud-based Web Applications} \cite{b9}, the authors explore an automatic method of generating a CI/CD configuration file, instead of writing their .gitlab-ci.yml file manually, which is error-prone. This ensures correctness of the CI/CD pipelines and prevents faults. Generation promotes the reuse of code instead of the usual copy/paste of code to similar projects which causes problems.

In another paper \textit{An Advanced DevOps Environment for Microservice-based Applications} \cite{b10}, the authors discuss providing reusable CI/CD templates for micro-service applications, which allows developers to deploy projects with no effort. These templates offer developers a low barrier to entry and doesn't requirer that they have knowledge of the details of the underlying pipeline. The improvement of the user experience when using templates increased the adoption of DevOps by simplifying the DevOps process. 

In a research thesis, titled \textit{Automation and Improvement of Software Development with GitLab CI Pipelines}, it was found that "Using templates simplifies the setup process by allowing developers to quickly  incorporate predefined configurations for common tasks such as building, testing,  and deploying applications. This results in reduced complexity, faster  onboarding for new team members, and a more consistent pipeline structure  across projects." \cite{b11}

In the master thesis titled \textit{Bridging Theory and Practice: Insights into Practical Implementations of Security Practices in Secure DevOps and CI/CD Environments} \cite{b12}, interviewees who are IT experts in different industry sectors have mentioned that there are many benefits to setting up templates and reusing them: they minimizing redundancy, reduce the amount work required when launching new applications, templates can go through a rigorous evaluation process which would be difficult to achieve if every pipeline was duplicated, there is a reduced potential of configuration errors, it leads to more efficient pipeline management, centrally managed templates have changes that reflect immediately across all instances, ensuring consistency across all pipelines and a consistent level of security. 

In the conference paper \textit{Configuration Tool for CI/CD Pipelines and React Web Apps}, the authors employ a YAML generator which analyzes the project files in order to generate their CI/CD pipelines for React web applications. Combining Lean, DevOps and Agile approaches, they increase efficiency through automation. This helped them overcome difficulties of the initial GitLab CI pipeline configuration and improved consistency. Managing complex React projects and ensuring consistency in CI/CD pipelines can be hard. The authors consider that building CI/CD pipelines is a repetitive task and is a waste of significant resources of time. Automated CI/CD generation tools were found to be potential solutions to manual work and human error, thus decreasing development time, accelerating the SDLC and improving deployment efficiency. \cite{b13}

In the Article \textit{Method for Continuous Integration and Deployment Using a Pipeline Generator for Agile Software Projects}, the authors used a CI/CD pipeline generator which generates pipelines from templates. Pipeline configuration files generated from the generator are ephemeral and can be configured to access extra jobs via a custom jobs configuration file. Because the steps to generate a pipeline are created instantly, deploying an application from scratch can take as low as 10 minutes. They've determined that this approach makes code duplication minimal and reduces financial costs. \cite{b15}

Although research on automatic CI/CD pipeline generation is scarce, the existing research points to a need and the presented solutions have shown a great potential in fulfilling this need.

\section{Conclusion}

In conclusion, this article has explored the concept of automatic pipeline provisioning within the context of Continuous Integration (CI) pipelines in software engineering projects. Through the implementation of automatic pipeline provisioning, organizations can streamline the process of creating and maintaining CI pipelines, reducing the burden on developers and ensuring consistency across projects. By leveraging templates, GUI-driven CI provisioning, fully automatic CI provisioning and/or Internal Developer Portals, companies can automate the generation of CI pipelines, saving time and effort while also improving the overall quality and reliability of their software development processes.

Case studies, such as the one presented with Company A, highlight the significant time and resource savings that can be achieved through automatic pipeline provisioning. While there may be initial investment required in setting up the automation infrastructure, the long-term benefits in terms of productivity and efficiency make it a worthwhile endeavor for organizations seeking to optimize their DevOps practices and improve their time-to-market.

In summary, automatic pipeline provisioning offers a promising solution to the challenges faced in managing CI pipelines, providing a more agile and scalable approach to software development in the era of DevOps.


\end{document}